\documentclass[preprint,prb,showpacs,preprintnumbers,amsmath,amssymb,floats]{revtex4}
\usepackage{graphicx}
\usepackage{dcolumn}
\usepackage{bm}
\usepackage{verbatim}

\newcommand{\KAoneprime}{$K_{A_1'}$}
\newcommand{\GamLO}{$\Gamma_{E_{2g}}$}

\newcommand{\EK}{$E_{K}$}
\newcommand{\EXtwo}{$E_{X_{2}}$}
\newcommand{\EXone}{$E_{X_{1}}$}
\newcommand{\Eoneone}{$E_{11}$}
\newcommand{\Etwotwo}{$E_{22}$}
\newcommand{\PLE}{$E_{11}^{(PLE)}$}
\newcommand{\PL}{$E_{11}^{(PL)}$}

\newcommand{\xone}{$X_1$}
\newcommand{\xtwo}{$X_2$}

\def\LevelDiagram{\begin{figure} [!t] \centering
\includegraphics[width=3.2in]{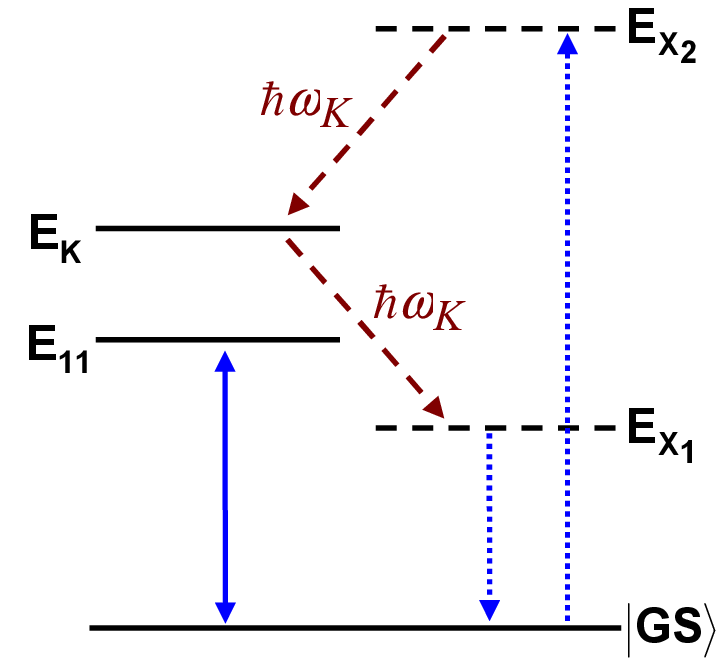} \caption{(Color online) Energy level diagram of the CNT excitonic states and phonon sidebands.  The solid blue double arrow indicates optical transitions involving the bright exciton.  The blue dotted arrows denote absorptive and emissive transitions involving the sidebands \xone\ and \xtwo\ of the dark exciton at \EK.  $\hbar\omega_K$ is the energy of a \KAoneprime\ phonon.}\label{LevelDiagram}
\end{figure}}

\def\PLMaps{\begin{figure} [!t] \centering
\includegraphics[width=3.2in]{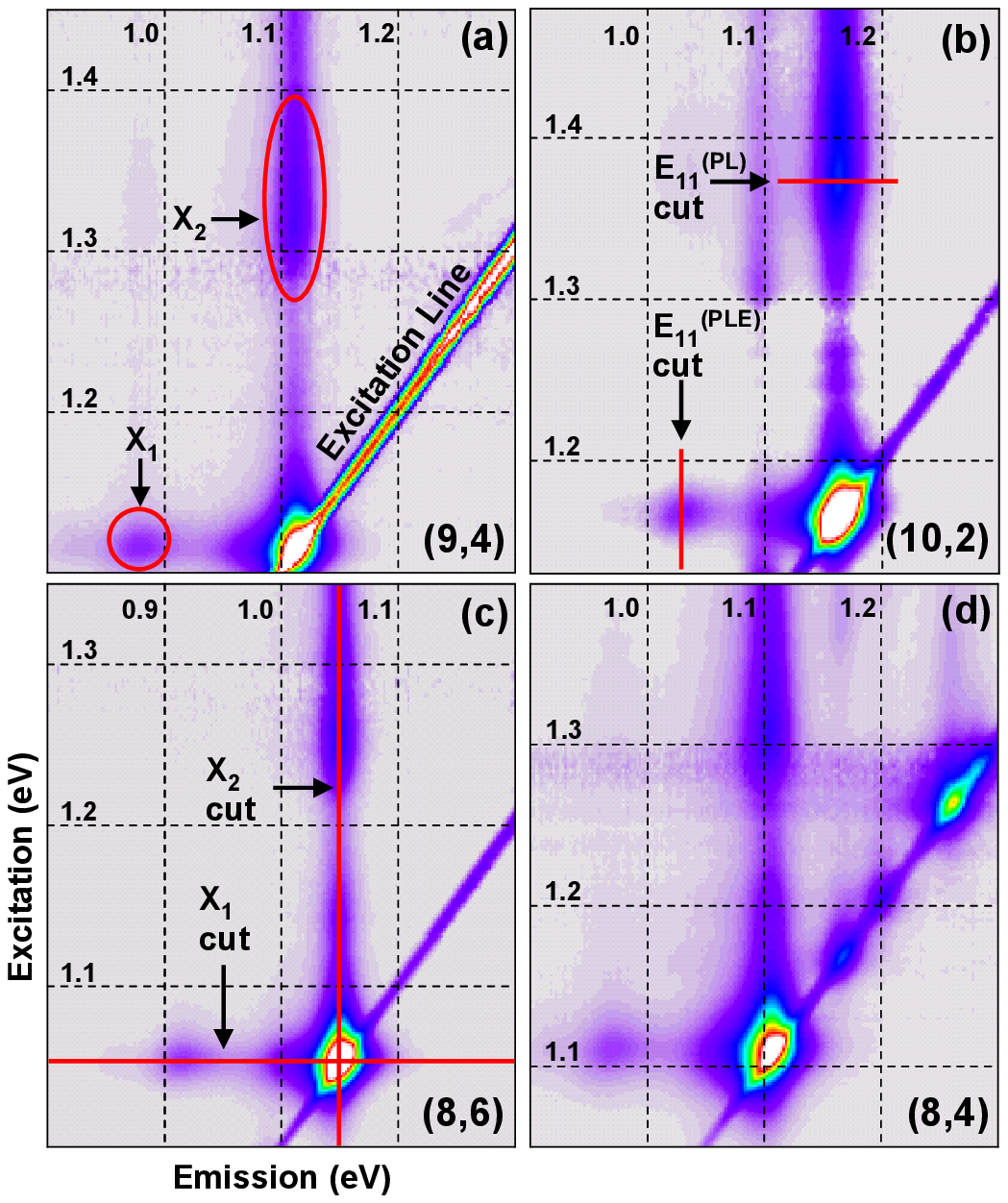} \caption{(Color online) PL maps of (a) $(9,4)$, (b) $(10,2)$, (c) $(8,6)$, and (d) $(8,4)$ CNTs.  The excitation line runs from upper right to bottom left, and the \Eoneone\ peak of the dominant species shows as the brightest spot along the excitation line.  Dashed lines indicate energies in eV, as labeled.}\label{PLMaps}
\end{figure}}

\def\OnRes{\begin{figure} [!t] \centering
\includegraphics[width=3.2in]{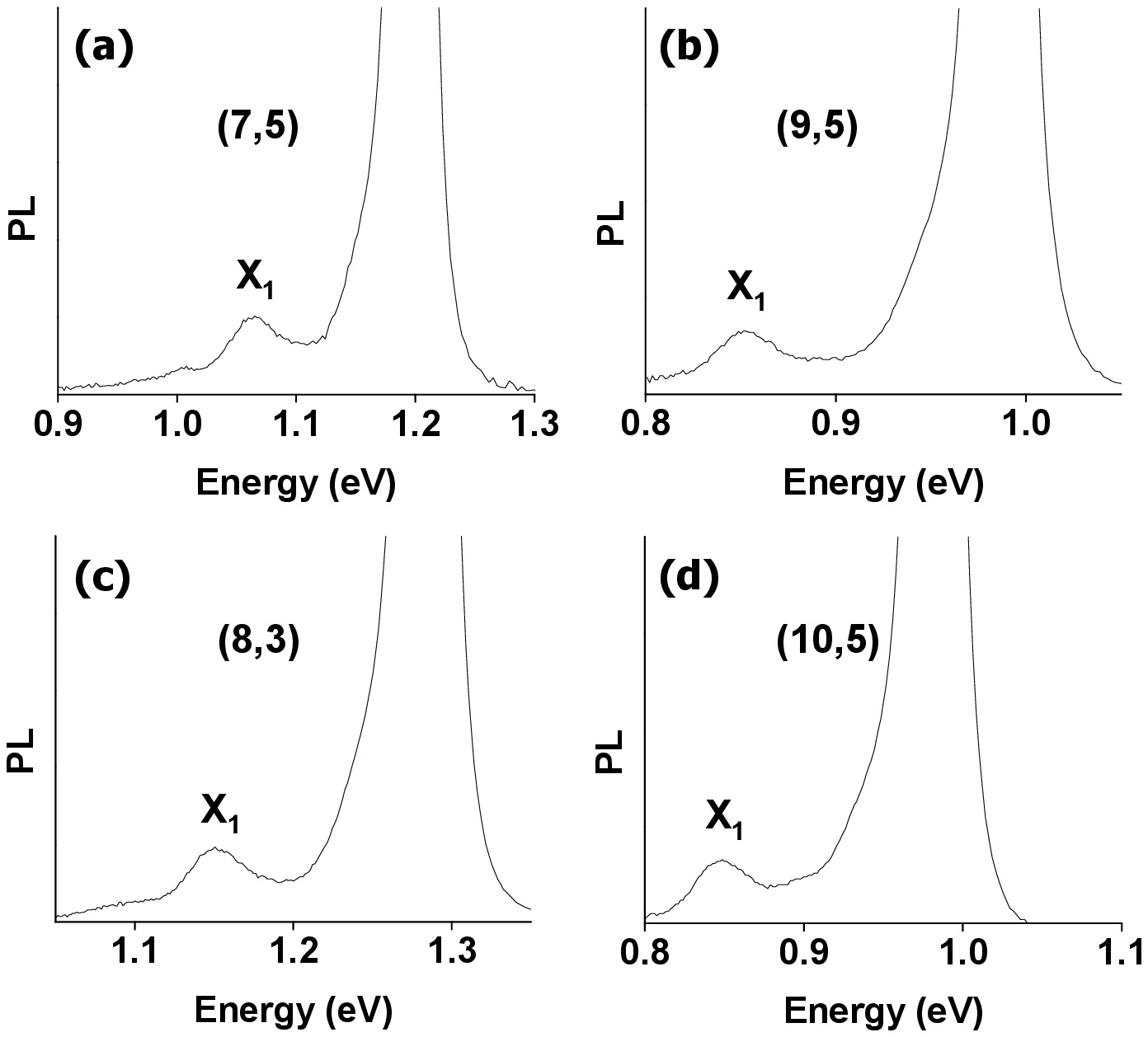} \caption{On-resonance PL spectra for (a) $(7,5)$, (b) $(9,5)$ (c) $(8,3)$, and (d) $(10,5)$ CNTs.}\label{OnRes}
\end{figure}}

\def\PLEspec{\begin{figure} [!t] \centering
\includegraphics[width=3.2in]{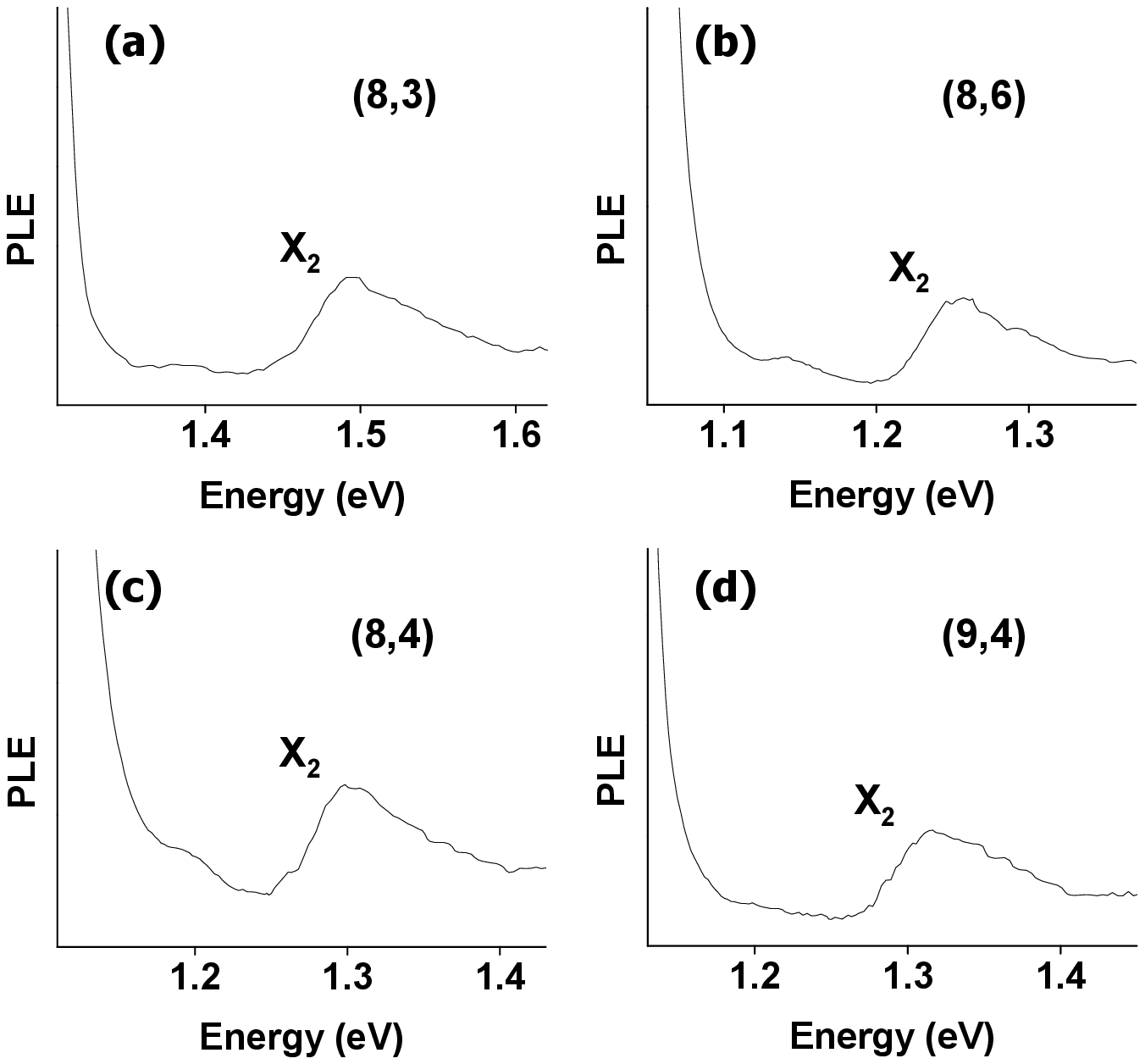} \caption{PLE spectra for (a) (8,3), (b) (8,6) (c) (8,4), and (d) (9,4) CNTs.}\label{PLEspec}
\end{figure}}

\def\DetEK{\begin{figure} [!t] \centering
\includegraphics[width=3.2in]{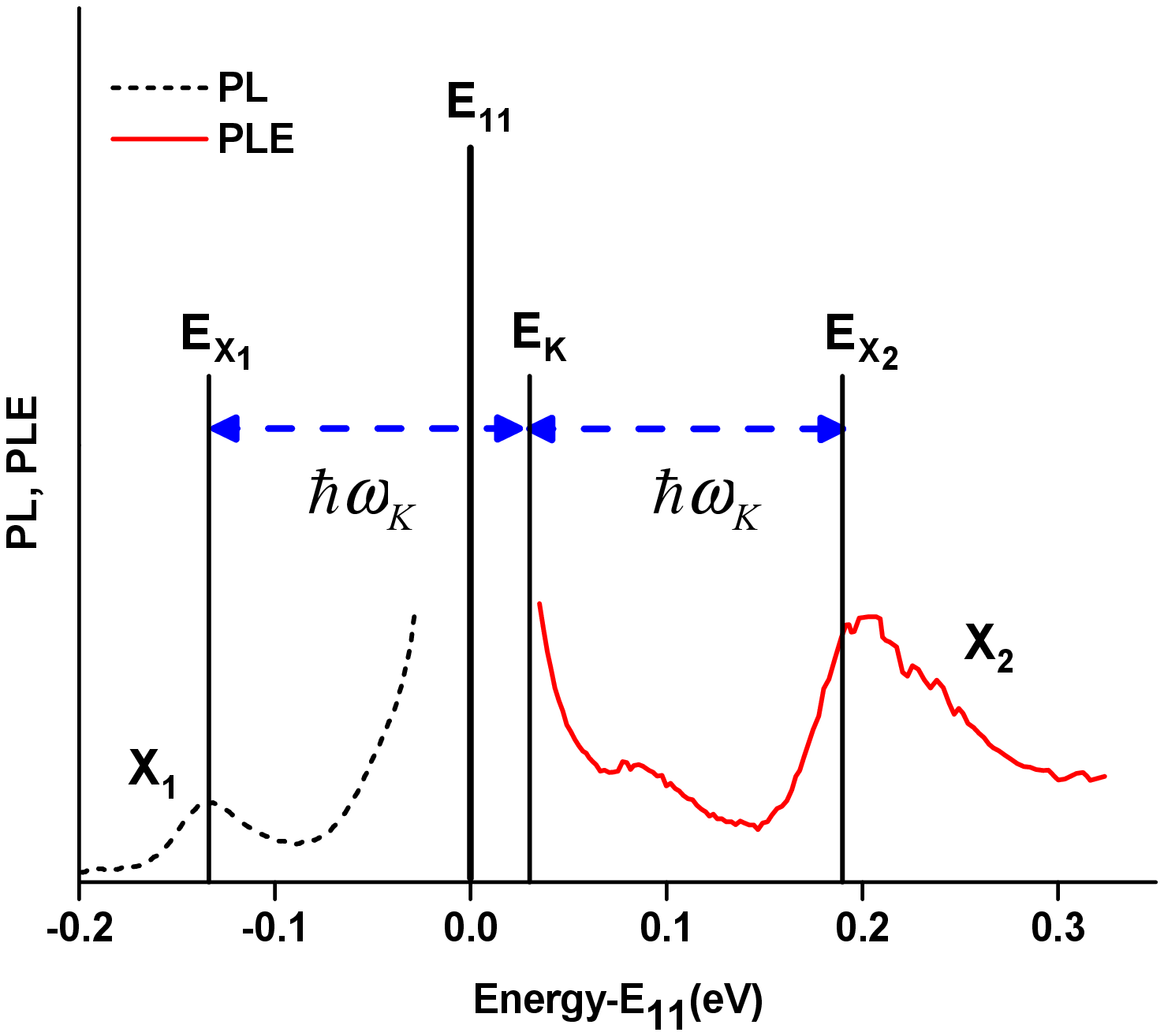} \caption{(Color online) PL (black, dashed) and PLE (red, solid) spectra of $(8,6)$ CNTs plotted on an energy axis relative to \Eoneone.  Vertical lines label \EXone, \EXtwo, \Eoneone, and \EK.  The blue dashed double arrowed lines label the energetic spacing of the sidebands from \EK, and $\hbar\omega_K$ is the energy of a \KAoneprime\ phonon.  Note that the peak of \xtwo\ is offset from \EXtwo\, as discussed in the text.}\label{DetEK}
\end{figure}}

\def\thexpLineX2{\begin{figure} [!t] \centering
\includegraphics[width=3.2in]{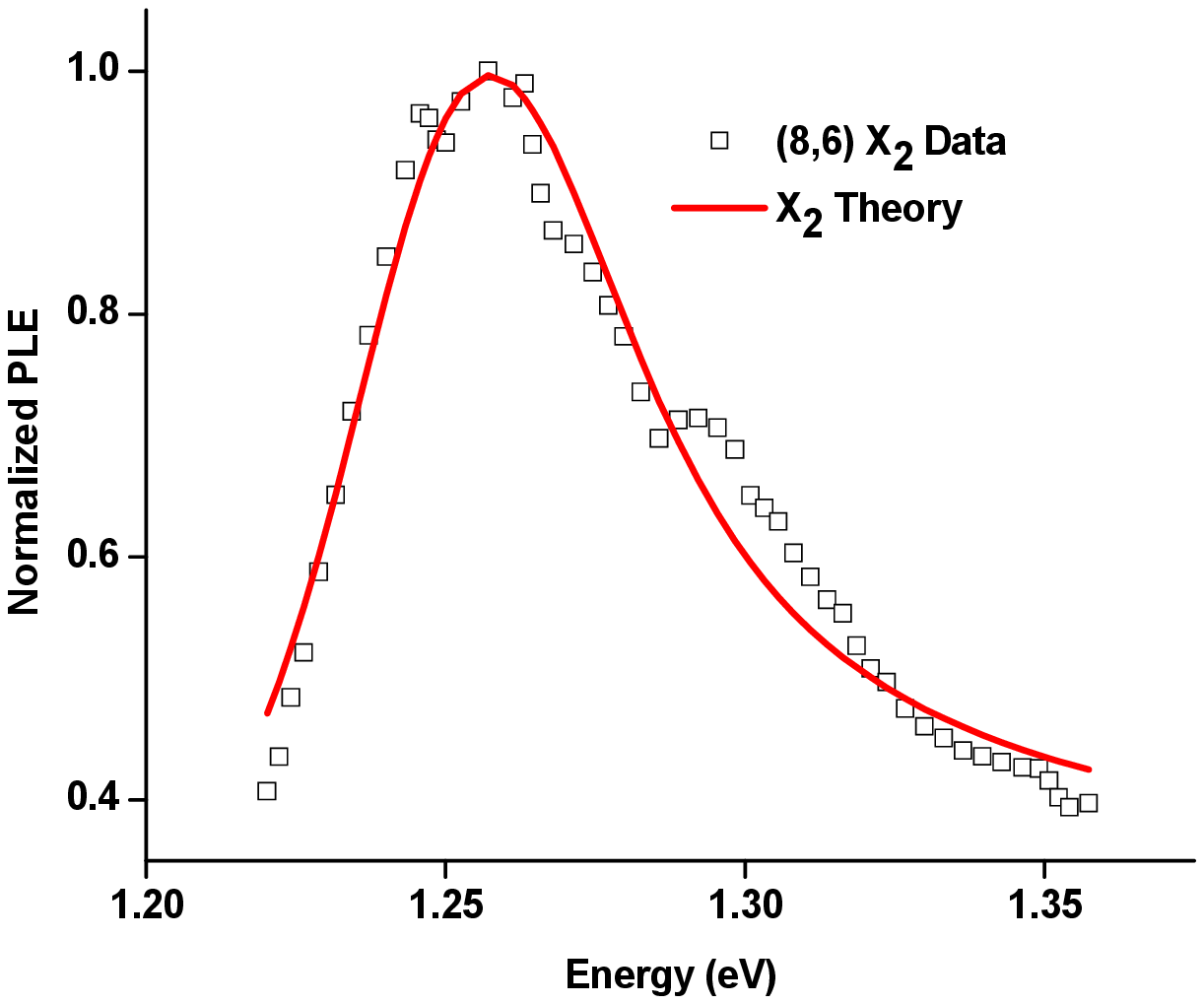} \caption{(Color online) Best fit of \xtwo\ for $(8,6)$ CNTs with the theoretical \xtwo\ lineshape described in the text.}\label{thexpLineX2}
\end{figure}}

\def\EKKatauraDataCapaz{\begin{figure} [!t] \centering
\includegraphics[width=3.2in]{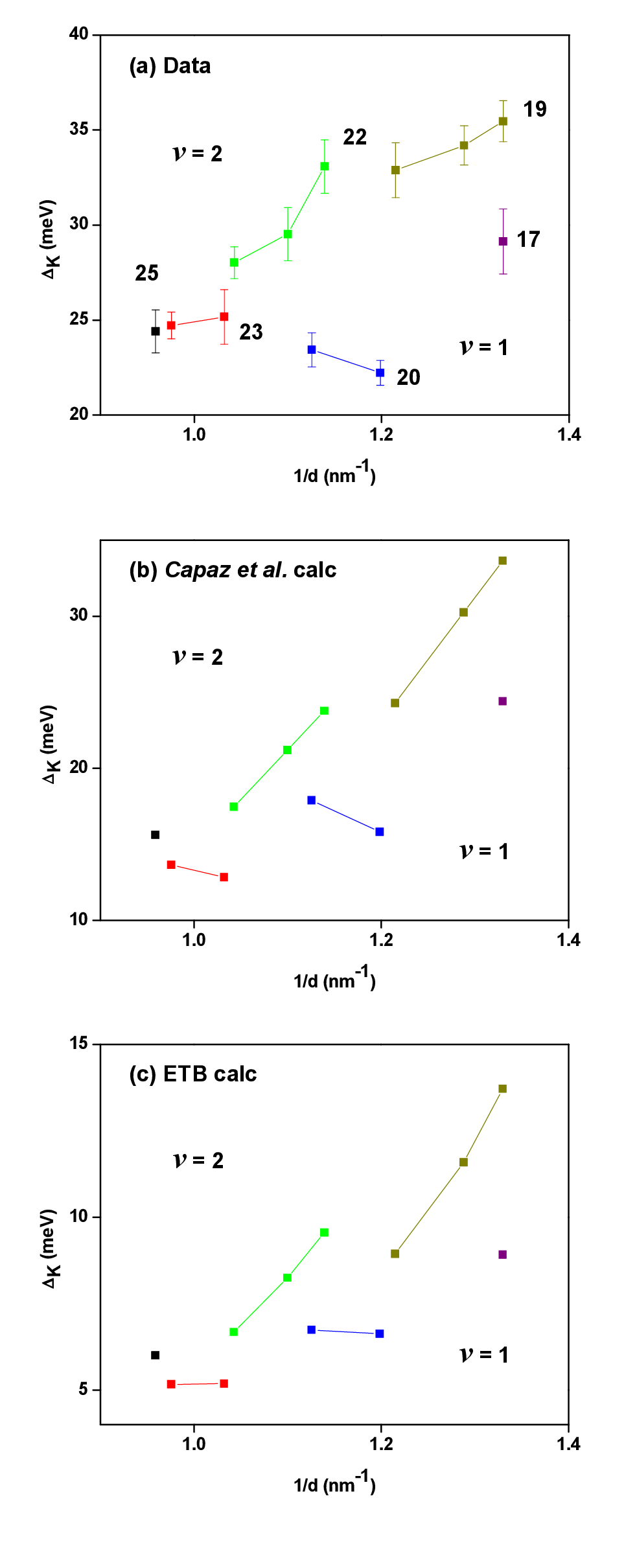} \caption{(Color online) $\Delta_K$ versus 1/d for (a) experimental data, (b) {\it ab inito} predictions from Ref. \onlinecite{PRB74(121401)} and (c) our ETB calculation.  Lines connect families sharing the same value of $2n+m$ indicated in (a).}\label{EK_KatauraDataCapaz}
\end{figure}}

\def\Exchange{\begin{figure} [!t] \centering
\includegraphics[width=3.2in]{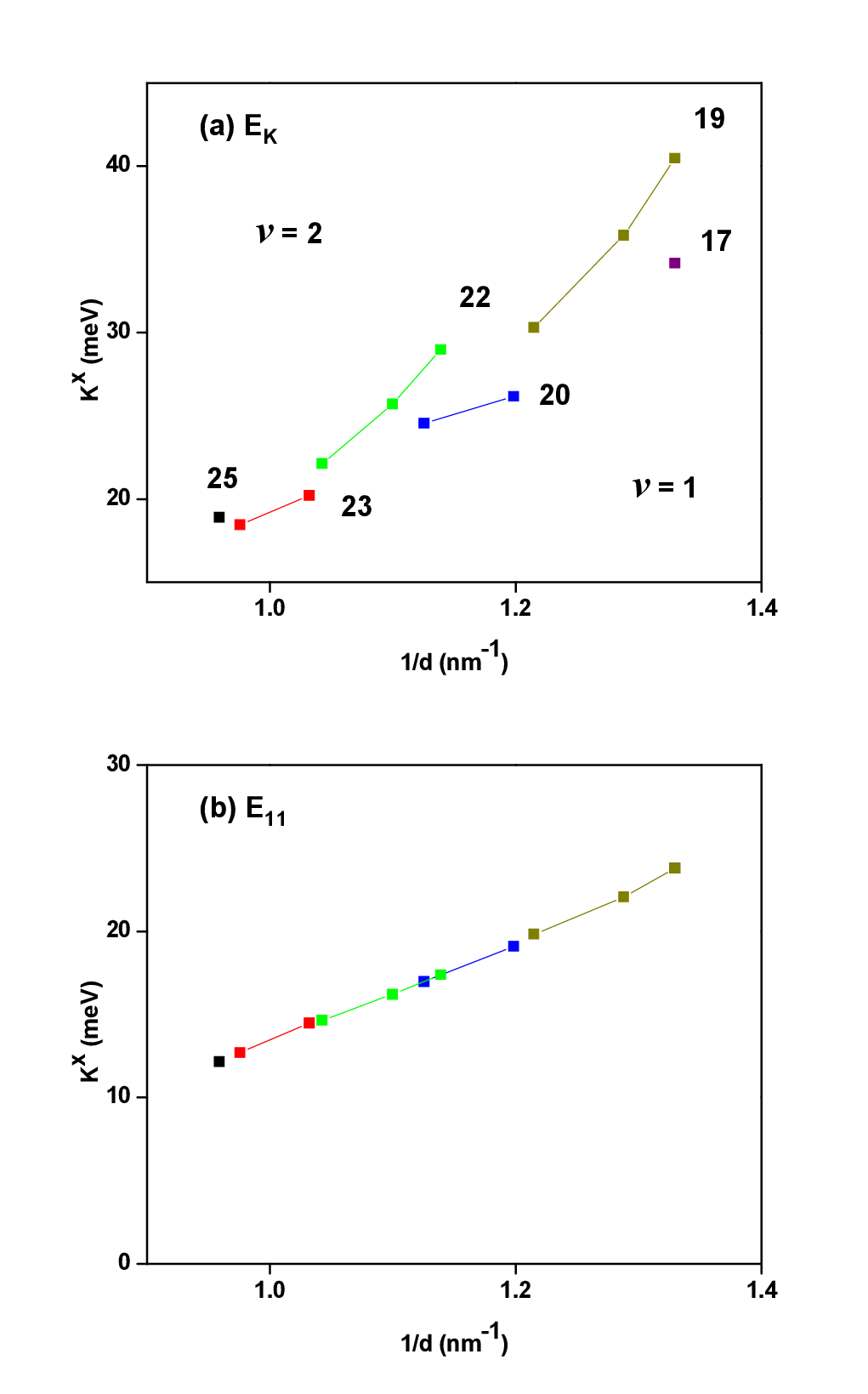} \caption{(Color online) Calculated exchange energy of (a) \EK\ and (b) \Eoneone\ versus 1/d. Lines connect families sharing the same value of $2n+m$ indicated in (a).}\label{Exchange}
\end{figure}}

\def\Scaling{\begin{figure} [t] \centering
\includegraphics[width=3.2in]{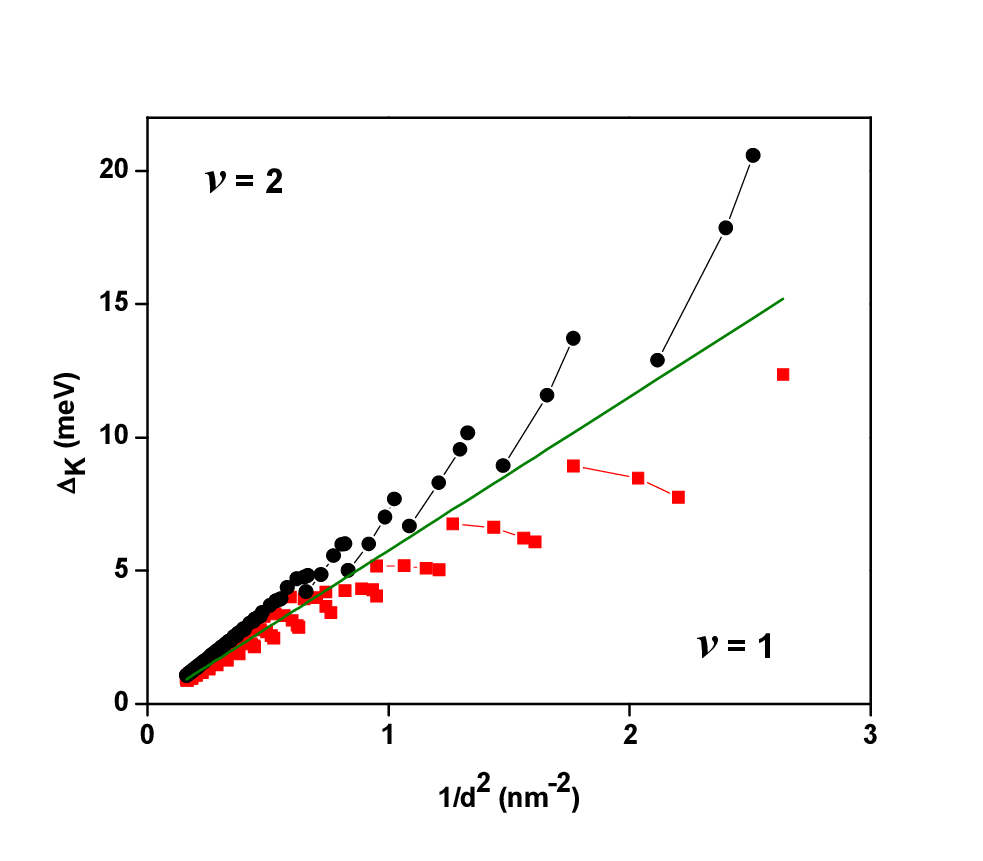} \caption{(Color online) $\Delta_K$ versus 1/d$^2$ calculated by our ETB model.  Red(black) squares(circles) are $\nu=1$(2) CNTs.  Lines connect families sharing the same value of $2n+m$, and the green line is a 1/d$^2$ fit.}\label{Scaling}
\end{figure}}

\def\SidebandRatio{\begin{figure} [!t] \centering
\includegraphics[width=3.2in]{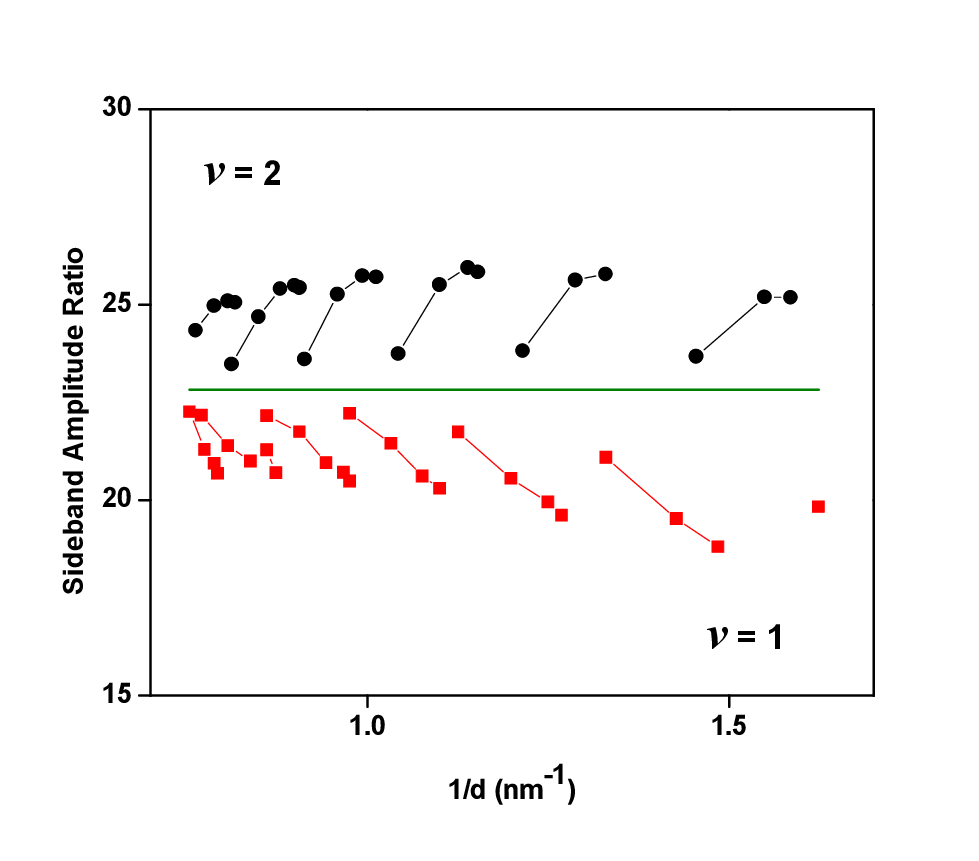} \caption{(Color online) Calculated ratio of \KAoneprime\ and \GamLO\ phonon contributions to \xtwo.  Red(black) squares(circles) are $\nu$=1(2) CNTs. Lines connect families sharing the same value of $2n+m$.  The green line labels an average value of 22.8.}\label{SidebandRatio}
\end{figure}}

\def\X1X2relE11{\begin{figure} [t] \centering
\includegraphics[width=3.2in]{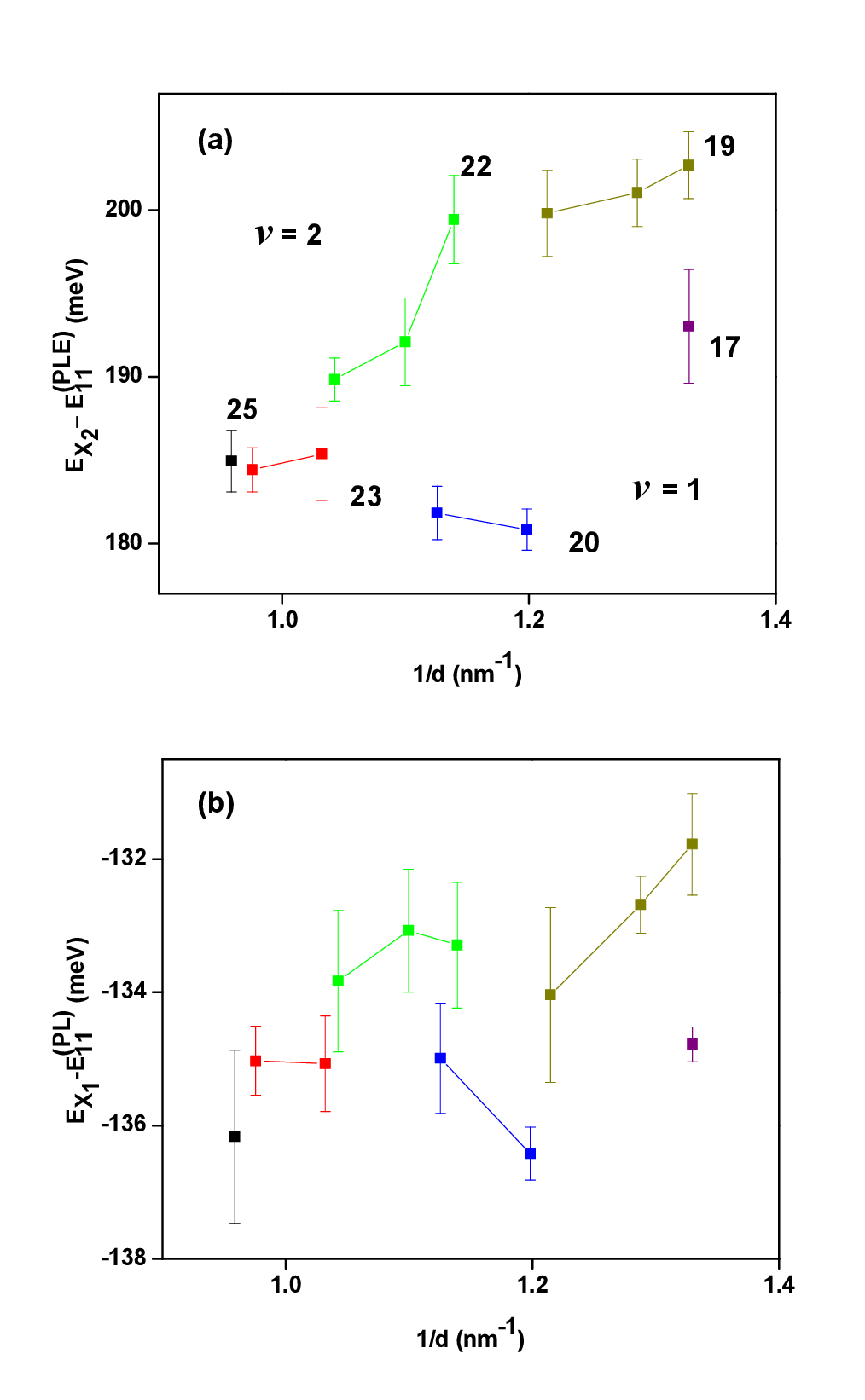} \caption{(Color online) Experimental data for (a) \EXtwo-\PLE\ and (b) \EXone-\PL\ versus 1/d.  Lines connect families sharing the same value of $2n+m$ indicated in (a).}\label{X1X2relE11}
\end{figure}}

\def\EKX2dataETBtheory{\begin{figure} [!t] \centering
\includegraphics[width=3.2in]{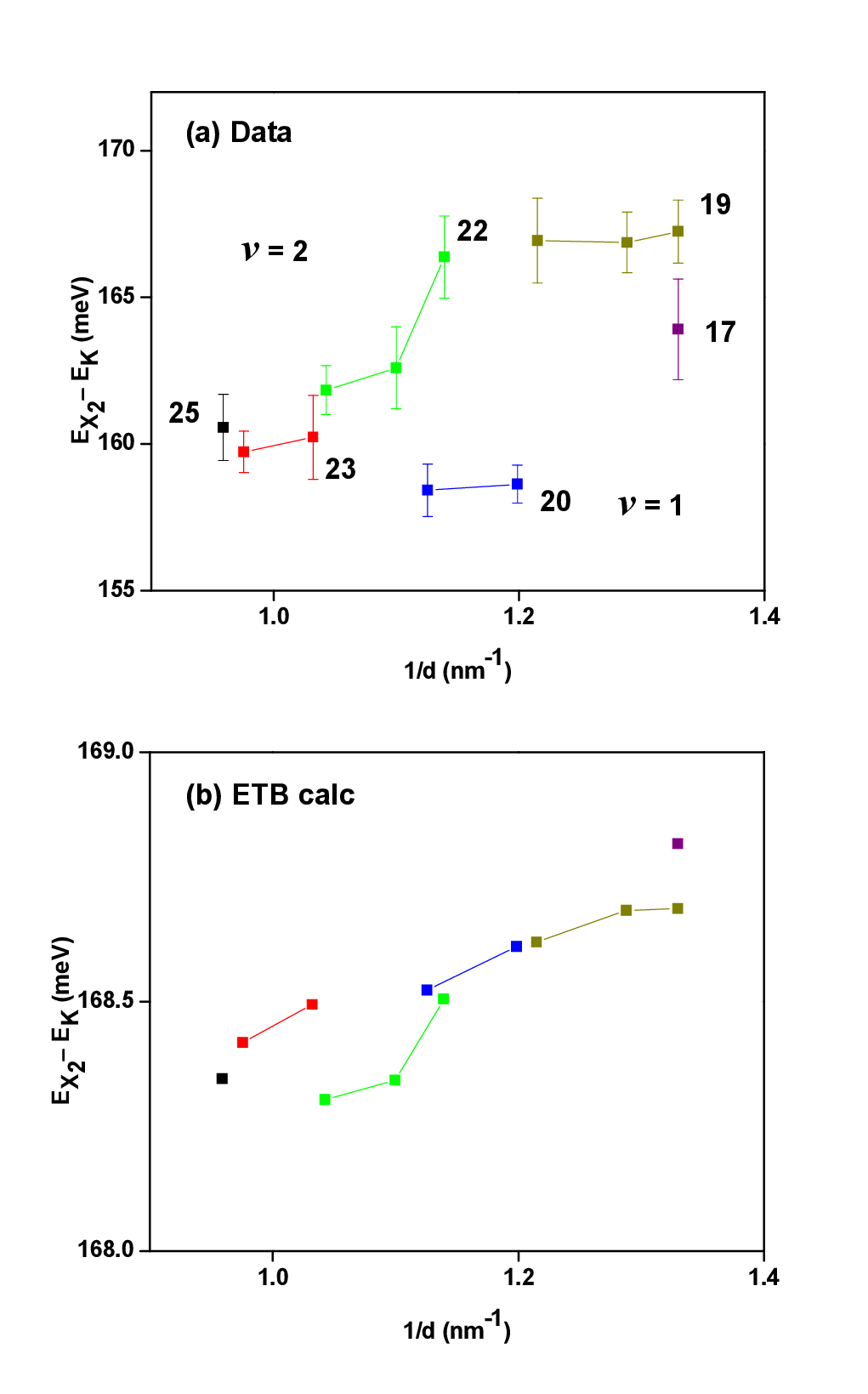} \caption{(Color online) \EXtwo$-$\EK\ versus 1/d for (a) experimental data and (b) our ETB calculation.  \EXtwo$-$\EK\ corresponds to half the sideband splitting, and nominally gives the energy of the \KAoneprime\ phonons. Lines connect families sharing the same value of $2n+m$ indicated in (a).}\label{EKX2dataETBtheory}
\end{figure}}

\def\DataTable{\begin{center}
\begin{table*}[!b]
\centering
\begin{tabular}{c|c|c|c|c|c}
(n,m) &\PLE\ & \PL\ & \EXone\ & \EXtwo\ & $\Delta_K$ \\
\hline
$(6,5)$ & 1.258 $\pm$ 0.0000 & 1.253 $\pm$ 0.0001 & 1.118 $\pm$ 0.0002 & 1.451 $\pm$ 0.0034 & 0.0291 $\pm$ 0.0034 \\
$(7,5)$ & 1.200 $\pm$ 0.0002 & 1.197 $\pm$ 0.0001 & 1.063 $\pm$ 0.0013 & 1.400 $\pm$ 0.0026 & 0.0329 $\pm$ 0.0029 \\
$(7,6)$ & 1.099 $\pm$ 0.0004 & 1.095 $\pm$ 0.0001 & 0.960 $\pm$ 0.0008 & 1.281 $\pm$ 0.0016 & 0.0234 $\pm$ 0.0018 \\
$(8,3)$ & 1.286 $\pm$ 0.0001 & 1.283 $\pm$ 0.0001 & 1.151 $\pm$ 0.0004 & 1.487 $\pm$ 0.0020 & 0.0342 $\pm$ 0.0021 \\
$(8,4)$ & 1.110 $\pm$ 0.0001 & 1.105 $\pm$ 0.0001 & 0.968 $\pm$ 0.0004 & 1.291 $\pm$ 0.0012 & 0.0222 $\pm$ 0.0013 \\
$(8,6)$ & 1.054 $\pm$ 0.0001 & 1.049 $\pm$ 0.0002 & 0.916 $\pm$ 0.0010 & 1.244 $\pm$ 0.0013 & 0.0280 $\pm$ 0.0017 \\
$(8,7)$ & 0.970 $\pm$ 0.0001 & 0.969 $\pm$ 0.0002 & 0.834 $\pm$ 0.0005 & 1.154 $\pm$ 0.0013 & 0.0247 $\pm$ 0.0014 \\
$(9,1)$ & 1.343 $\pm$ 0.0005 & 1.340 $\pm$ 0.0001 & 1.208 $\pm$ 0.0007 & 1.546 $\pm$ 0.0019 & 0.0355 $\pm$ 0.0021 \\
$(9,4)$ & 1.115 $\pm$ 0.0002 & 1.111 $\pm$ 0.0003 & 0.978 $\pm$ 0.0009 & 1.307 $\pm$ 0.0026 & 0.0295 $\pm$ 0.0028 \\
$(9,5)$ & 0.989 $\pm$ 0.0000 & 0.987 $\pm$ 0.0001 & 0.852 $\pm$ 0.0007 & 1.174 $\pm$ 0.0029 & 0.0252 $\pm$ 0.0029 \\
$(10,2)$ & 1.167 $\pm$ 0.0003 & 1.162 $\pm$ 0.0000 & 1.029 $\pm$ 0.0009 & 1.366 $\pm$ 0.0026 & 0.0331 $\pm$ 0.0028 \\
$(10,5)$ & 0.983 $\pm$ 0.0000 & 0.982 $\pm$ 0.0002 & 0.846 $\pm$ 0.0013 & 1.168 $\pm$ 0.0018 & 0.0244 $\pm$ 0.0023 \\ [1ex]
\end{tabular}
\caption{Measured values of \PLE, \PL, \EXone, \EXtwo, and $\Delta_K$ in eV.}
\label{DataTable}
\end{table*}
\end{center}}

\preprint{APS/123-QED}

\begin{document}

\title{Chirality Dependence of the $K$-Momentum Dark Excitons in Carbon Nanotubes}

\author{P. M. Vora$^1$, X. Tu$^2$, E. J. Mele$^1$, M. Zheng$^2$, J. M. Kikkawa$^{1,*}$}
\affiliation{$^1$Department of Physics and Astronomy, University
of Pennsylvania, Philadelphia, Pennsylvania 19104, USA}

\affiliation{$^2$DuPont Central Research and Development, Experimental Station, Wilmington, Delaware 19880, USA}

\date{\today}

\begin{abstract}

Using a collection of twelve semiconducting carbon nanotube samples, each highly enriched in a single chirality, we study the chirality dependence of the $K$-momentum dark singlet exciton using phonon sideband optical spectroscopy.  Measurements of bright absorptive and emissive sidebands of this finite momentum exciton identify its energy as 20 - 38 meV above the bright singlet exciton, a separation that exhibits systematic dependencies on tube diameter, $2n+m$ family, and semiconducting type.  We present calculations that explain how chiral angle dependence in this energy separation relates to the the Coulomb exchange interaction, and elaborate the dominance of the \KAoneprime\ phonon sidebands over the zone-center phonon sidebands over a wide range of chiralities.  The Kataura plot arising from these data is qualitatively well described by theory, but the energy separation between the sidebands shows a larger chiral dependence than predicted.   This latter observation may indicate a larger dispersion for the associated phonon near the $K$ point than expected from finite distance force modeling.

\end{abstract}

\maketitle

\section{Introduction}
Semiconducting carbon nanotubes (CNTs) have attracted widespread interest due to their unique electronic structure.  Their quasi one-dimensional nature makes electron-electron interactions important, leading to the formation of strongly bound exciton states.  The absorption and emission spectra of CNTs are well known to be dominated by excitons, allowing for a direct study of many body physics in a quasi-1D regime.\cite{Science308(838),PRL90(207401),JPSJ66(1066)}  The lowest energy manifold of excitons is comprised of sixteen states:  four singlets and twelve triplets.  Only one of these excitons is bright (optically active).\cite{PRL93(157402),PRL92(257402),NanoLett5(749)}  The remaining fifteen excitons are dark (optically inactive) and have remained largely unstudied for this reason.

Experimental investigations of dark excitons initially focused on the zero momentum dark singlet exciton, which can be brightened by the application of an external magnetic field.\cite{PRL101(087402),NanoLett7(1851),PRL96(016406)}  Lattice defects induced by intense pulsed laser excitation can also brighten otherwise dark exciton states.\cite{NanoLett9(2010),PRB81(033401)}  Beyond the zero momentum dark singlet, there remain two time reversal degenerate dark singlet excitons with center-of-mass momenta near the $K$ and $K'$ points of the graphene Brillouin zone (BZ).  While these $K$-momentum dark singlet excitons (at energy \EK) cannot be directly photoexcited due to momentum conservation, they can be excited in combination with a $K$-momentum phonon.  This process is enhanced by the strong exciton-phonon coupling near the $K$ points of the graphene BZ, resulting in the appearance of pairs of bright phonon sidebands located {\it asymmetrically} about the bright exciton.\cite{PRL94(027402),PRL101(157401)}   Previously, several of us used these sideband pairs to study the $K$-momentum exciton in a purified $(n,m)=(6,5)$ CNT sample, where $(n,m)$ denotes the CNT wrapping vector.\cite{PRL101(157401)}  The amplitude, shape and separation of the sidebands was found to be in good agreement with a theory modeling them as bright sidebands of the dark $K$-momentum singlet exciton.   Nevertheless, the inferred separation of \EK\ from the bright singlet (\Eoneone) exceeded theoretical predictions, and the sidebands have had other interpretations in the literature.\cite{PRB74(035415),PRL95(247401),NanoLett7(1203),PRB72(241402),PRL99(237402)} Hence, a more comprehensive study over a wider range of chiralities is called for.

Figure \ref{LevelDiagram} illustrates our approach in using sideband spectroscopy to identify \EK.  Phonons of $A_1'$ symmetry at the $K'$ ($K$) point (iTO branch) create dynamic $\sqrt{3}\times\sqrt{3}$ intervalley scattering potentials that couple excitons at the $K$ ($K'$) point to excitons at the zone center.  (For a visualization of this phonon mode, the reader is referred to Figure 7 of Ref. \onlinecite{PRB76(035439)}, which illustrates the atomic displacements labeled as $K_1$.) The associated hybridization creates bright phonon sidebands symmetrically about \EK.  We denote the lower energy emissive sideband \xone, and the upper energy absorptive sideband \xtwo.  One can think of \xtwo\ as creating a phonon-exciton pair, each at the zone boundary but with zero combined momentum, a transition made possible through their mixing with the bright exciton.   Similarly, \xone\ corresponds to an emissive process in which the $K$-momentum exciton creates a photon but conserves momentum by emitting a phonon.  As indicated in the level diagram, the energetic separation between \xone\ and \xtwo\ is essentially twice the \KAoneprime\ phonon energy, and measurement of both sidebands permits an interpolation of \EK.  Two recent works have studied one of these sidebands, \xone, over a range of chiralities but did not reveal a systematic chirality dependence.\cite{PRB79(195407),PRB81(033401)}  In contrast, here sideband {\it pairs} are simultaneously identified in twelve highly purified single chirality CNT suspensions permitting the determination of \EK.  Our high resolution study finds that both \EXone\ and \EXtwo\ depend on chirality, and that qualitative agreement with theory emerges in the pattern of energies produced by a comprehensive study of \EK. These findings lead to a physical understanding of the processes that dominate this dark-bright splitting.  Specifically, whereas the chiral family behavior of the bright exciton emerges predominantly from the single quasiparticle energies, the dark-bright splitting measured here reveals chiral behavior arising from the many-particle exchange interaction.

Our study shows that  \xone\ and \xtwo\ appear asymmetrically about the bright exciton for each species, and exhibit systematic chiral dependences.   Kataura plots of $\Delta_K=$ \EK$-$\Eoneone\ show $2n+m$ family behavior, semiconducting type dependence, and an overall scaling with diameter.  A recent theoretical study\cite{PRB74(121401)} and our own extended tight binding (ETB) calculations predict behavior consistent with these results, and a $1/d^2$ scaling of $\Delta_K$ in the large diameter limit.  Because the \xtwo\ absorptive sideband has been interpreted as a \GamLO\ (longitudinal optical mode at the $\Gamma$-point) phonon sideband of the bright exciton,\cite{PRB74(035415),PRL95(247401),NanoLett7(1203)} we calculate the relative contributions of the \GamLO\ and \KAoneprime\ phonons to \xtwo\ over a large diameter range, and find the \KAoneprime\ sideband dominates by more than an order of magnitude in all cases.  This prediction, along with the rather strong chiral dependence of the experimental \xtwo\ sideband energy, strongly discourages against further interpretation of \xtwo\ as arising from zone-center phonons.   The energetic spacing between \xone\ and \xtwo\ should be given by two \KAoneprime\ phonon energies.  Experimentally, we note that this separation shows a larger chiral dependence than expected, suggesting the \KAoneprime\ phonons may be more dispersive near the $K$ point than current finite distance force models predict.

\section{Experimental Methods}

Highly purified single chirality dispersions of $(6,5)$, $(7,5)$, $(7,6)$, $(8,3)$, $(8,4)$, $(8,6)$, $(8,7)$, $(9,1)$, $(9,4)$, $(9,5)$, $(10,2)$, and $(10,5)$ CNTs are prepared following the method of Ref. \onlinecite{Nature460(250)}.  The output of a 150 W Xenon arc lamp passes through the first two 0.5 meter stages of a triple spectrometer in a bandpass configuration.   The resulting 10 nm wide excitation spectrum contains very low spectral content outside the bandpass window, and is further filtered to remove second order diffractive components.  Aqueous suspensions are excited in precision quartz cuvettes with excitation and emission beams at 90 degrees. Photoluminescence (PL) is collected into the third stage of our triple spectrometer, and detected with a TE-cooled silicon CCD (visible) and a Li $\mbox{N}_2$ cooled InGaAs array (near-infrared).  An additional 0.5 meter spectrometer and TE-cooled InGaAs array are used as a power meter.   To ensure proper averaging over the spatial profile of the excitation beam, a beamsplitter samples part of the excitation and focuses it onto a diffuser first before collection into the power meter.  All elements of our setup are calibrated for spectral response with a NIST traceable light source and all reported spectra are power normalized.  The \Eoneone\ absorption resonance is frequently used in our analysis, and is typically determined using PL excitation profiles discussed in Section III.  We find that optical absorption spectra give the same value of \Eoneone\ and are used as a supplemental measurement.  Absorption data collected for this purpose are obtained using a tungsten lamp and the aforementioned triple 0.5 meter spectrometer and detectors.

By combining PL emission spectra taken at different excitation energies, we assemble two dimensional excitation-emission PL maps.  In our study the range of excitation energies usually passes directly through the \Eoneone\ resonance, known as resonant PL spectroscopy.  Typically, broadband excitation scatter would limit the efficacy of resonant PL spectroscopy for these samples by obscuring interesting features close to the excitation energy.  Here, the low spectral background of our excitation source makes resonant PL spectroscopy an effective tool for studying \xone\ and \xtwo\ (Figure \ref{PLMaps}).

\section{Data Analysis}\label{dataanalysis}

Theoretical calculation methods described in Refs. \onlinecite{PRL94(027402)}, \onlinecite{SSC130(657)} and \onlinecite{PRB75(035407)} and adopted by several of us in Ref.~\onlinecite{PRL101(157401)} are frequently used in our analysis.  We elaborate these methods here for completeness.  First, we obtain a single-particle basis of states using an extended tight binding (ETB) model with a nearest neighbor overlap integral of 2.7 eV.  We then use the Bethe-Salpeter equation (BSE) to calculate exciton energies $E^{S}_\mathbf{q}$ and wavefunctions,
\begin{equation}
|\Psi^S_q>=\sum_{\substack{\mathbf{k}}} A_\mathbf{k,q}^S c^\dagger_{\mathbf{k+q},c} c_{\mathbf{k},v}|\textbf{GS}>,
\end{equation}
\noindent following the methods of Ref.~\onlinecite{PRB75(035407)}. Here, $A^S_\mathbf{k,q}$ are the exciton wavefunction coefficients for and electron-hole pair with center-of-mass momentum $\mathbf{q}$ and relative momenta $\mathbf{k+q/2}$, $S=0,1,...$ indexes the ground and excited states of the exciton, and $c^\dagger_{\mathbf{k},i}$ is an electron creation operator for wavevector $\mathbf{k}$ in the conduction ($i=c$) or valence ($i=v$) band. Our simulation uses 20,000 atoms and an environmental dielectric constant of 2.7.\cite{PRL101(157401)} We compute the phonon eigenspectrum using a symmetry-adapted valence force constant model\cite{SSC130(657)} up to fifth nearest neighbors.\cite{PRB76(035439)}  We then calculate the electron-phonon couplings $\mathrm{M}_\mathbf{k,q}^\mu$ to first order in the lattice deformations,~\cite{PRB22(904)} with a nearest-neighbor coupling constant of $5.3$~eV/\AA.~\cite{PRL94(027402)} Here, $\mu=1,2,...,6$ is the phonon band index. The electron-phonon coupling is then used to compute the {\it exciton}-phonon coupling,\cite{PRL94(027402)}
\begin{equation}\label{Beqn}
B_{\mathbf{q}\mu}^{SS'} = \sum_\mathbf{k} \mathrm{M}_\mathbf{k,q}^\mu A^{S'*}_\mathbf{k,q} (A^S_{\mathbf{k},0}+A^S_{\mathbf{k+q},0}).
\end{equation}
\noindent As described in Ref.~\onlinecite{PRL94(027402)}, mixing of the bright exciton with excitons at finite momentum $\mathbf{q}$ and phonons at $-\mathbf{q}$ gives an unbroadened absorptive sideband lineshape of the form,
\begin{equation}\label{intensity}
I(\omega ) \propto \delta (E_{11}-\hbar \omega)+
\sum_{\mathbf{q},\mu,S'} \frac{ |B_{\mathbf{q}\mu}^{0S'}|^2 }{(E^{S'}_\mathbf{q} + \hbar \omega_{-\mathbf{q}}^\mu -E_{11})^2}  \delta ( E^{S'}_\mathbf{q} + \hbar \omega_{-\mathbf{q}}^\mu - \hbar \omega ).
\end{equation}
\noindent In the above expression, $\omega_{-\mathbf{q}}^\mu$ is the phonon eigenfrequency at wavevector $-\mathbf{q}$, and $E_{11}$ is the bright exciton energy.~\cite{PRL94(027402)} The second term represents the \xtwo\ sideband line shape, and in modeling our data we convolve this form with a Gaussian to represent inhomogeneous broadening due to potential fluctuations.  Our calculation also produces theoretical values of energies \EXtwo, \EK, and \Eoneone.

We measure the energy of the $K$-momentum exciton through phonon sideband optical spectroscopy, in which the average energy of \xone\ and \xtwo\ gives \EK, and these features are both separated from \EK\ by the energy of one \KAoneprime\ phonon.\cite{PRL101(157401)}  Representative PL maps, resonant PL spectra, and photoluminescence excitation (PLE) spectra from our measurements are shown in Figures \ref{PLMaps}, \ref{OnRes}, and \ref{PLEspec}, respectively.  PL maps taken on all samples show \xone\ and \xtwo\ features resonantly aligned with \Eoneone\ absorption or emission, respectively, associating these transitions with the dominant chirality of the sample.  A subset of these maps appears in Fig. \ref{PLMaps}, with \xone\ and \xtwo\ identified by arrows in Fig. \ref{PLMaps}(a). The energetic location of these sidebands relative to \Eoneone\ is always asymmetric, in accord with Ref. \onlinecite{PRL101(157401)} and placing \EK\ consistently above \Eoneone.   Overlapping transitions from unintended chiralities are suppressed by using samples highly enriched in a single chirality.  Further discrimination against features associated with other chiralities is obtained by analyzing vertical or horizontal line cuts through the data that pass through the \Eoneone\ resonance (Fig. \ref{PLMaps}(c)).  Emission(excitation) spectra containing the \xone(\xtwo) sideband are thereby obtained at the excitation(emission) energy that places \Eoneone\ and \xone(\xtwo) into resonance, a subset of which are shown in Fig. \ref{OnRes} (Fig. \ref{PLEspec}).   The location of these line cuts is indicated in Figure \ref{PLMaps}(c), with the spectrum resulting from a \xone\ cut heretofore referred to as an ``on-resonance'' PL spectrum, and that arising from a \xtwo\ cut denoted a photoluminescence excitation (PLE) spectrum.   PLE is essentially an absorbance spectroscopy, but gives superior background rejection compared to conventional absorbance, as the latter cannot discriminate among contributions from different chiralities.

Note that here we are interested in the relative energy separation of \xone\ and \xtwo, but \xone\ contains a Stokes shift since emission resonances in CNTs generally shift a few meV lower in energy than their absorptive counterparts.  A relative energy scheme, as shown in Figure \ref{DetEK}, then becomes quite useful.   In this plot, \xone\ and \xtwo\ are placed on a common energy axis and measured relative to \PL\ and \PLE, respectively.\cite{PRL101(157401)}  Here, \PL\ (\PLE) is the energetic location of \Eoneone\ that appears in the PL (PLE) spectra and is summarized in Table I.  To avoid problems discriminating PL emission from excitation scatter at the peak of the \Eoneone\ resonance, we actually determine \PL\ and \PLE\ by taking line cuts across the relatively weaker \xtwo\ and \xone\ features at locations indicated in Figure \ref{PLMaps}(b) and performing Gaussian or Lorentzian fits.  We have found that absorbance provides the same values of \PLE\ as these line cuts, and because chiral discrimination is not an issue in determining the location of the strong \Eoneone\ feature, we use whichever method gives the least error in determining \PLE.

We remark that the PLE spectra shown in Figures \ref{PLEspec} and \ref{DetEK} all show evidence of a small feature $\sim$90 meV above \PLE.  Our theoretical calculations predict the presence of a phonon sideband of \EK\ at this energy, arising from $K$-momentum oTA and oTO phonon modes.  However, our modeling suggests the intensity of this feature relative to \xtwo\ is only $\sim$1/200 and would be unobservable in these measurements.  Hence, the identification of this minor sideband remains an open question to the community.

Some analysis is required to accurately determine energetic locations of \xone\ and \xtwo, denoted \EXone\ and \EXtwo, respectively.  Theory suggests these features should lie symmetrically about \EK, but their line shapes differ significantly.   \xtwo\ is an absorptive process with a continuum of states available above a lower energy threshold, and the resulting asymmetric line shape follows closely the density of available exciton-phonon final states.\cite{PRL101(157401)}  \xone\ is an emissive process, so that thermal relaxation drives the excited population toward the lowest energy excitons having center-of-mass momenta closest to the $K$ point.  The resulting more symmetric line shape thus resembles that of \Eoneone.  These spectral profiles influence our determination of \EXone\ and \EXtwo.  In the case of  \EXone, \xone\ happens to lie on the tail of the \Eoneone\ emission.  We thus subtract a linear background before fitting \xone\ with a Lorentzian to obtain its central value.

The determination of \EXtwo\ includes an offset that is related to line broadening of an asymmetric line shape.  To understand this relationship, imagine an asymmetric unbroadened profile resembling $\Theta(E-E_1)\text{exp}(-E/E_0)$, where $\Theta$ is a step function.  If we apply a Gaussian broadening to this function the resulting line shape will peak above $E_1$.   The profile of \xtwo\ before broadening is qualitatively similar to this function, with an absorbance threshold at \EXtwo$\equiv$\EK$+\hbar\omega_K$, where $\hbar\omega_K$ is the energy of the \KAoneprime\ phonon closest to the $K$ point.   We find that 30 meV Gaussian broadening produces a $\sim$8 meV blue shift of the peak of \xtwo\ relative to \EXtwo, which is rather insensitive to chirality.  A broadening of 42 meV gives a larger 11 meV shift, and is used here because it produces the best fit to \xtwo\ (Fig. \ref{thexpLineX2}).  To determine \EXtwo, we fit \xtwo\ to determine the location of its peak and then subtract the aforementioned shift, explicitly computed for that particular chirality.  The \xtwo\ fitting line shape is used only to determine the peak position, is generic to all chiralities, and is calculated for a $(6,5)$ CNT.   Ultimately, these data are used to locate the position of \EK, and in that regard we emphasize two important points.  First, variations in the shift with chirality are small and even if we do not account for it the family behavior of \EK\ remains the same.   Second, systematic errors arising from the choice of broadening result in a shift of the inferred value of \EK\ that is constant across all chiralities and of order 1 meV, again leaving all reported family behavior unchanged.

Defining the splitting energy of the $K$-momentum exciton relative to the bright exciton as $\Delta_K\equiv $~\EK$-$\Eoneone, we obtain $\Delta_K =($\EXone$+$\EXtwo$-2$\Eoneone$)/2$ (Fig. \ref{DetEK}).  Table \ref{DataTable} summarizes \PL, \PLE, \EXone, \EXtwo, $\Delta_K$, and their non-systematic errors.  Data for $(6,5)$ CNTs was taken from Ref. \onlinecite{PRL101(157401)} and reanalyzed following the above procedure.  This results in a somewhat different, but more accurate, value of $\Delta_K$ than previously reported.\cite{PRL101(157401)}

\section{Discussion and Summary}

Experimental results for $\Delta_K$ versus inverse tube diameter $(1/d)$ are shown in Figure \ref{EK_KatauraDataCapaz}(a).  Predictions for $\Delta_K$ by an \textit{ab initio} parameterized variational calculation\cite{PRB74(121401)} and our ETB model are shown in Figures \ref{EK_KatauraDataCapaz}(b) and \ref{EK_KatauraDataCapaz}(c).  In the experimental data, we observe a grouping of $\Delta_K$ by $2n+m$ families and a dependence upon chiral index (a so-called ``Kataura plot'').  The chiral index $\nu=(n-m)\mbox{ mod }3$ typically classifies two distinct groups of semiconducting family behaviors.  In our case $\nu = 1(2)$ type semiconducting tubes have a $\Delta_K$ that fans downwards (upwards) in energy.  The theoretical results of Ref. \onlinecite{PRB74(121401)} and our ETB calculations show the same semiconducting and family groupings seen in the data.  The ETB model underestimates the value of $\Delta_K$, while the variational calculation is in much better agreement.\cite{PRB74(121401)}

$\Delta_K$ is the difference between the binding energies of \EK\ and \Eoneone, and thus a direct measure of electron-electron interactions in semiconducting CNTs.  The physics governing many particle interactions is contained within the interaction kernel, $\mathcal{K}=2\mathcal{K}^x-\mathcal{K}^d$, of the Bethe-Salpeter equation.\cite{PRB75(035407),PRB62(4927)}  $\mathcal{K}^d(\mathcal{K}^x)$ is the direct(exchange) interaction.  In Appendix A, we study the properties of the interaction kernel for \Eoneone\ and \EK.  The direct interaction is usually responsible for exciton formation\cite{PRB62(4927)} except in unusual circumstances,\cite{PRL92(077402)} and has only a weak dependence on chiral angle for \Eoneone\ and \EK.  Indeed the direct energy is nearly identical in the two excitons and thus has little effect on $\Delta_K$.  The exchange interaction, however, differs significantly between the two excitons.  For \EK, the exchange interaction involves elements of the Coulomb interaction in the vicinity of the $K$ point.  This makes the associated exchange energy (Figure \ref{Exchange}(a)) relatively sensitive to the threefold lattice anisotropy.\cite{PRB61(2981)}  In contrast, the exchange interaction for \Eoneone\ involves only small momentum properties of the interaction kernel, which are less sensitive to chiral angle variations (Figure \ref{Exchange}(b)). Therefore, the family fan-out behavior of $\Delta_K$ derives from chiral variation of the exchange energy in \EK, not \Eoneone.

The dependence of exciton splittings on $d$ is currently an open question.  One {\it ab initio} study predicted a $1/d^2$ dependence for $\Delta_K$,\cite{PRB74(121401)} and another using the same computational methods employed here has predicted a $1/d$ splitting between zero momentum singlets and triplets.\cite{NanoLett5(2495)}  We note that the diameter ranges for these two calculations are non-overlapping, with the former focusing on small diameter tubes and the latter seeking to establish a large diameter scaling limit.   Here we find the diameter scaling of $\Delta_K$ cannot be uniquely determined from our data set, but we are nevertheless able to establish a $1/d^2$ behavior within our calculations for CNTs with diameters lying between 0.6 nm and 2.5 nm (Figure \ref{Scaling}).  This scaling behavior is identical to that of the exchange energy responsible for $1/d^2$ behavior in Ref. \onlinecite{PRB74(121401)}. Although the authors of Ref.~\onlinecite{NanoLett5(2495)} did not study $\Delta_K$, the difference in power law scaling within the exciton manifold predicted by the {\it same} methods is interesting.

The nature of \xtwo\ remains a controversial topic in the CNT photophysics community.  It is not uncommon to see \xtwo\ interpreted as a \GamLO\ phonon sideband of the bright \Eoneone\ exciton, despite recent theoretical\cite{PRL94(027402),PRL101(157401)} and experimental\cite{PRL101(157401)} works indicating that for certain chiralities the dominant contribution comes from \KAoneprime\ phonons.  Here we apply our ETB calculation to all CNTs with diameters ranging between 0.6 nm and 1.33 nm and extract the contributions of the \GamLO\ and \KAoneprime\ phonon modes to the exciton-phonon sideband.  The findings strongly support our assignment, as the \KAoneprime\ component dominates over that of \GamLO\ by more than an order of magnitude for all chiralities (Figure \ref{SidebandRatio}).

The authors of Ref.~\onlinecite{PRB79(195407)} recently studied \xone\ using PL maps excited near the \Etwotwo\ resonance in $(6,5)$, $(7,5)$, and $(10,5)$ CNT samples.  Their measurements indicate that \EXone\ lies $\sim$140 meV below \PL\ and does not depend on tube diameter or chiral angle.\cite{PRB79(195407)}  Even more recently, the authors of Ref. \onlinecite{PRB81(033401)} observed \xone\ in individual CNTs through temperature dependent PL spectroscopy, reporting \EXone\ values $\sim$130 meV below \PL\ that exhibit little diameter dependence.  In contrast to these studies, we examine both \xone\ \textit{and} \xtwo, allowing for the observation of chiral behavior in \EXone, \EXtwo\ and, most importantly, \EK, which cannot be determined from \xone\ alone.  Katuara plots of our results for \EXtwo, \EXone\ relative to \PLE, \PL\ are summarized in Figure \ref{X1X2relE11}(a), \ref{X1X2relE11}(b) respectively.  Both \EXone\ and \EXtwo\ show similar chiral family behavior, but the variation in \EXone\ is approximately 1/5 that found in \EXtwo\ and was therefore too small to be resolved by the aforementioned studies.  There are two implications of this behavior.  First, the sideband spacing approximates two \KAoneprime\ phonon energies and depends on CNT chirality.  Some dependence of the \KAoneprime\ phonon energy on chirality is generically expected from the proximity of the nearest allowed k-vectors to the $K$ point, an effect proportional to the phonon dispersion.  However, our calculation shows the expected variation due to this effect is much smaller than seen in the data.   Figure \ref{EKX2dataETBtheory} shows that ETB predicts \EXtwo$-$\EK$ (\sim\hbar\omega_K$) to have a variation $\sim$0.5 meV within the chiralities studied, whereas our data finds $\sim$8 meV.  This difference might reflect a limitation of our phonon calculation.  Graphene displays Kohn anomalies at the $K$ and $\Gamma$ points in the iTO and LO graphene phonon branches.\cite{SSC143(47),PRL92(075501),PRL93(185503),PRB76(233407)}  Kohn anomalies lead to a strong softening of the phonon energies, resulting in cusps in the dispersion of the optical phonon branch at the $\Gamma$ and $K$ points.  Short range force constant models such as those used here cannot describe this non-analytic behavior, and therefore generically underestimate the phonon dispersion.\cite{PRL93(185503)}  For CNTs, quantization in the circumferential direction significantly enhances the Kohn anomaly in metallic species,\cite{PRB75(035427)} but in semiconducting CNTs the electronic gap provides a long distance cutoff to the interatomic force constants and prevents a true Kohn anomaly.  Our data suggest the intriguing possibility that there may nevertheless be some enhancement of the iTO phonon dispersion near the $K$ point in semiconducting CNTs.  A second implication of the data in Figure \ref{X1X2relE11} is that the chiral variations in $\Delta_K$ mirror those in $\hbar\omega_K$.  We have not yet identified a fundamental reason why this should be the case.

\section{Conclusion}

In this manuscript, we present the chirality dependent behavior of the $K$-momentum dark singlet exciton in twelve highly purified single chirality CNT samples.  These observations complete the characterization of singlet excitons in CNTs.  $\Delta_K$ exhibits $2n+m$ family behavior, chiral index dependence, and diameter dependence in agreement with ETB calculations and a theoretical study.\cite{PRB74(121401)}  This behavior is found to originate from the exchange energy of \EK.  We calculate the scaling behavior of $\Delta_K$ and suggest a $1/d^{2}$ law in agreement with Ref. \onlinecite{PRB74(121401)}.  The \KAoneprime\ phonon energy varies strongly with chirality, in disagreement with our ETB calculations.  The origin of this behavior is unclear, but may be related to the Kohn anomalies of graphene.  Finally, ETB calculations indicate that \xtwo\ arises from \KAoneprime\ phonons. We hope that these results can guide the community in further efforts to understand the behavior of dark excitons and their role in the optical properties of carbon nanotubes.

JMK and PMV acknowledge the support of NSF DMR-0907266 and MRSEC DMR-05-20020.  EJM was supported by the Department of Energy under grant DE-FG02-ER45118.

\appendix

\section{Interaction Kernel}

We follow the methods outlined in Ref. \onlinecite{PRB75(035407)} to understand the effect of chirality on the interaction kernel,\cite{PRB75(035407)}

\begin{align}
\mathcal{K}(\mathbf{k_c^\prime}\mathbf{ k_v^\prime, k_c k_v})=&2 \mathcal{K}^x(\mathbf{k_c^\prime k_v^\prime,k_c k_v})\\ \nonumber
&-\mathcal{K}^d(\mathbf{k_c^\prime k_v^\prime,k_c k_v}).
\end{align}

\noindent$\mathcal{K}^d$ is the direct interaction, typically responsible for exciton formation,\cite{PRB62(4927)} $\mathcal{K}^x$ is the exchange interaction, and $\mathbf{k_c,\text{ }k_v}$ denote wavevectors in the conduction and valence bands.  The two interaction terms may be expressed as\cite{PRB75(035407)}
\begin{align}
\mathcal{K}^x(\mathbf{k_c^\prime}\mathbf{ k_v^\prime, k_c k_v})=&\int d\mathbf{r^\prime} d\mathbf{r} \psi_{\mathbf{k_c^\prime}}^* (\mathbf{r^\prime}) \psi_{\mathbf{k_v^\prime}} (\mathbf{r^\prime})\\ \nonumber
&\times v(\mathbf{r^\prime,r}) \psi_{\mathbf{k_c}}(\mathbf{r}) \psi_{\mathbf{k_v}}^*(\mathbf{r}),
\end{align}

\begin{align}
\mathcal{K}^d(\mathbf{k_c^\prime}\mathbf{ k_v^\prime, k_c k_v})=&\int d\mathbf{r^\prime} d\mathbf{r} \psi_{\mathbf{k_c^\prime}}^* (\mathbf{r^\prime}) \psi_{\mathbf{k_c}} (\mathbf{r^\prime})\\ \nonumber
&\times w(\mathbf{r^\prime,r}) \psi_{\mathbf{k_v^\prime}}(\mathbf{r}) \psi_{\mathbf{k_v}}^*(\mathbf{r}),
\end{align} where $v$ and $w$ are the bare and screened Coulomb interactions, and $\psi$ are the single quasiparticle wavefunctions.\cite{PRB75(035407)}

Consider \Eoneone\ created with zero center-of-mass momentum and finite relative momentum near the $K$ point, $\mathbf{K+k}$.  The direct term for a momentum transfer $\mathbf{q}$ in the $\mathbf{q\to0}$ limit is,

\begin{align}
\lim_{\mathbf{q\to0}} \mathcal{K}^d(\mathbf{q}) \propto w(\mathbf{q\to0}) \int d \mathbf{r^\prime} &d \mathbf{r} |\psi_{\mathbf{K+k},c}(\mathbf{r^\prime})|^2 \\ \nonumber
&\times |\psi_{\mathbf{K+k},v}(\mathbf{r})|^2.
\end{align}

\noindent This interaction involves the Coulomb potential $w(\mathbf{q})$ for $|\mathbf{q}|a\ll1$, where $a$ is the lattice constant, and is therefore relatively insensitive to lattice anisotropy.

The exchange interaction,

\begin{align}
\mathcal{K}^x(\mathbf{q})=v(\mathbf{q}) &\int d\mathbf{r^\prime} d\mathbf{r} \psi_{\mathbf{K+k+q},c}^* (\mathbf{r^\prime}) \psi_{\mathbf{K+k+q},v} (\mathbf{r^\prime})\\ \nonumber
&\times e^{-\imath \mathbf{q} \cdot (\mathbf{r^\prime-r})}  \psi_{\mathbf{K+k},c}(\mathbf{r}) \psi_{\mathbf{K+k},v}^*(\mathbf{r}),
\end{align}

\noindent vanishes in the $\mathbf{q\to0}$ limit and at finite momentum provides a smaller contribution to the exciton binding energy.  The small momentum $|\mathbf{q}|a\ll1$ terms in the exchange are again insensitive to the lattice anisotropy.

The direct and exchange interactions for \Eoneone\ thus depend only weakly on chiral angle, implying that the observed behavior in $\Delta_K$ must originate from \EK.  The direct interaction for \EK\ involves scattering of a particle and hole within their respective valleys, in perfect analogy with the direct interaction for \Eoneone.  For \EK, the initial and final momenta are $\mathbf{k_c=K^\prime+k,\text{ }k_v=K-k}$ and $\mathbf{k_c^\prime=K^\prime+k+q,\text{ }k_v^\prime=K-k+q}$.   The direct interaction in the $\mathbf{q\to0}$ limit is

\begin{align}
\lim_{\mathbf{q\to0}} \mathcal{K}^d(\mathbf{q}) \propto w(\mathbf{q\to0})\int d\mathbf{r^\prime} &d\mathbf{r} |{\psi_{\mathbf{K^\prime+k},c}} (\mathbf{r^\prime})|^2\\ \nonumber
&\times |\psi_{\mathbf{K-k},v}(\mathbf{r})|^2.
\end{align}

\noindent The direct interaction for \EK\ is nearly identical to that of \Eoneone, insensitive to chiral angle for the same reasons.

The exchange interaction for \EK\ is very different from \Eoneone, since exchanging a particle and hole for an intervalley process requires a minimum momentum transfer of $\mathbf{2K}$.  All relevant Coulomb interaction matrix elements will be of order $\mathbf{K}$ and sensitive to the lattice anisotropy.  The exchange may be written as,

\begin{align}
\mathcal{K}^x(\mathbf{q})=&\int d\mathbf{r^\prime} d\mathbf{r} {\psi^*_{\mathbf{K^\prime+k+q},c}} (\mathbf{r^\prime}) \psi_{\mathbf{K-k+q},v} (\mathbf{r^\prime})\label{Kexchange} \\ \nonumber
&\times v(\mathbf{q+2K}) e^{-\imath (\mathbf{q+2K}) \cdot (\mathbf{r^\prime-r})}\notag \\ \nonumber
&\times \psi_{\mathbf{K^\prime+k},c}(\mathbf{r}) \psi^*_{\mathbf{K-k},v}(\mathbf{r}).\notag
\end{align}  In the $\mathbf{q\to0}$ limit, \eqref{Kexchange} becomes

\begin{align}
\lim_{\mathbf{q\to0}} \mathcal{K}^x(\mathbf{q}) \propto &\int d\mathbf{r^\prime} {\psi^*_{\mathbf{K^\prime+k},c}} (\mathbf{r^\prime}) \psi_{\mathbf{K-k},v} (\mathbf{r^\prime})\label{exchange1} \\ \nonumber
&\int d\mathbf{r} \psi_{\mathbf{K^\prime+k},c} (\mathbf{r}) \psi^*_{\mathbf{K-k},v} (\mathbf{r})\notag \\ \nonumber
&\times v(\mathbf{2K}) e^{-\imath (\mathbf{2K}) \cdot (\mathbf{r^\prime-r})}.\notag
\end{align}

Time reversal symmetry implies that $\psi_{\mathbf{K^\prime+k},c}(\mathbf{r})=\psi_{\mathbf{K-k},v}(\mathbf{r})$ up to a phase factor, allowing \eqref{exchange1} to be rewritten as,

\begin{align}
\lim_{\mathbf{q\to0}} \mathcal{K}^x(\mathbf{q}) \propto  &\int d\mathbf{r^\prime} |\psi_{\mathbf{K^\prime+k},c} (\mathbf{r^\prime})|^2\int d\mathbf{r} |\psi_{\mathbf{K^\prime+k},c} (\mathbf{r})|^2 \\ \nonumber
&\times v(\mathbf{2K})e^{-\imath (\mathbf{2K}) \cdot (\mathbf{r^\prime-r})}.
\end{align}

\noindent $v(\mathbf{2K})$ is responsible for the chiral angle dependence observed in $\Delta_K$.  Although $v(\mathbf{2K})$ is very small, the exchange interaction is proportional to the intervalley overlap integrals and produces a much larger anisotropic contribution than the other direct and exchange terms discussed here.  Note that $\mathbf{2K}$ is equivalent to $\mathbf{K^\prime}$ where lattice anisotropy effects on the potential are relatively large.

\noindent $^*$To whom communication should be addressed.  Email: kikkawa@physics.upenn.edu

\newpage
\LevelDiagram
\PLMaps
\OnRes
\PLEspec
\DetEK
\thexpLineX2
\EKKatauraDataCapaz
\Exchange
\Scaling
\SidebandRatio
\X1X2relE11
\EKX2dataETBtheory
\DataTable

\end{document}